\documentclass[aps,prd,showpacs,nofootinbib,preprintnumbers,twocolumn]{revtex4-1}
\usepackage[ansinew]{inputenc} 
\usepackage{graphicx,epsfig}
\usepackage{url}
\usepackage{color}
\usepackage[colorlinks=true,linkcolor=black,allcolors=red]{hyperref}
\hypersetup{colorlinks,linkcolor={blue},citecolor={red},urlcolor={red}}
\usepackage{bm}
\usepackage{dcolumn}
\usepackage{amsfonts,amssymb,amsmath}
\usepackage{psfrag}
\usepackage{subfigure}
\usepackage{latexsym}
\begin{document}
\title{Variation in the size of the Photon Sphere and Black Hole Shadow in the Modified Gravity}
\author{Ajay Kumar  Sharma}
\email{aksh.sharma2@gmail.com}
\affiliation{Department of Physics, University of Lucknow, Lucknow, 226007, India}
\author{Kanav Ganguly}
\email{kanav.ganguly@gmail.com}
\affiliation{Lassiter High School, 2601 Shallowford Road, Marietta, Georgia, 30066, United States}
\author{Murli Manohar Verma}
\email{sunilmmv@yahoo.com}
\affiliation{Department of Physics, University of Lucknow, Lucknow, 226007, India}
\date{\today}
\begin{abstract}
Black hole shadows are a widespread topic in astrophysics. This paper searches for an optical view of the black hole and the relationship between black hole shadow and photon sphere with curvature. We were inspired by the observations of Sagittarius $A^*$ and supermassive black hole M87$*$ through the Event Horizon Telescope.  In this paper, we have found the signature of the modified theory of gravity on the photon sphere and shadow. We considered $f(R)$ modified theory of gravity. Our finding is that the new scalar degree of freedom $F$ appears in the expressions of photon sphere and black hole shadow. Both are changed with $F$, and indeed, they depends on spacetime metric.
\end{abstract}

\maketitle
The first image of the shadow was captured in 2019 by Event Horizon Telescope (EHT) \cite{EventHorizonTelescope:2019ggy}. This is a groundbreaking observation black hole and used for studying the nature of black hole. The observational result of EHT is useful for testing the theory of gravity.


Till now, we do not have direct observational evidences about the dark matter particle. Another mysterious component of the  universe is dark energy, which responsible for the accelerated expansion of the universe. There are many methods and theories to solve both mysteries, dark matter and dark energy. The simplest way to tackle these problems by introducing a $\Lambda$ term for late time expansion \cite{Riess:1998fmf, Carroll:2000fy, Sahni:2002kh, Sahni:1999gb, Sahni:2015hbf} and Cold Dark Matter (CDM) as a candidate of dark matter \cite{Bertone:2004pz, Armendariz-Picon:2013jej, Ilic:2020onu, Perivolaropoulos:2021jda, Yang:2025ume}. There are many other way to explain them and one of them is the modification in the curvature. We used $f(R)$ models for accelerated expansion and large scale structure formation \cite{Sharma2022fiw, Sharma:2022tce} and dark matter \cite{Yadav:2018llv, Sharma:2019yix, Sharma:2020vex}. 

This article seeks to unravel the complexities surrounding black holes in the photon sphere \cite{snepppen2021divergent}, the very first layer of these bodies. These are the boundaries of the Black Hole Shadow (BHS).

On the other hand, shadow is like an illusion of four dimensions of space-time and it depends on the many factors like background metric, static or moving observer and its position etc. Recent picture taken by the Event Horizon Telescope (EHT) of the Sagittarius $A^*$ (Sgr A*, pronounced "sadge-ay-star") \cite{akiyama2022firstv1}.

The trajectory of light bends around the object because of high-density. If light bends so strongly that it moves on circular path because of a high density object like a black hole \cite{snepppen2021divergent}. These orbits of light are known as photon spheres. Dark spots in the middle are termed as black hole shadow. In 1966, Synge studied path of the photons for null geodesics \cite{synge1966escape} for Schwarzschild and non-rotating black hole.

In the past few years, many researchers have been investigated the black hole shadows. We know that the size of the black hole shadow is 2.4 times larger than the particle horizon \cite{Wang:2022kvg, Perlick2021CalculatingBH}. This introduces some concept about the illusion of black hole shadow. It can be explained with the help of location of the observer. That means BHS depends on observer's position, accretion disc that means matter around the horizon, rotation of black hole, and charged black hole \cite{takahashi2004shapes,li2020shadow, gralla2019black}. Shapes of BHS strongly depends on modified theory of gravity models. Black hole shadow can be strong feature of the BH that constraint model's parameters of the modified theory of gravity. 
\section{Observational Evidence}
The image of the M87$^*$ black hole taken by the Event Horizon Telescope (EHT) \cite{2024A&A681A79E} reveals that the angular radius \cite{EventHorizonTelescope:2019dse} $42 \pm 3$ micro arc-seconds ($\mu as$), and angular gravitational radius 3.8 $\pm$ 0.4 $\mu$ac, mass $M = (6.5\pm0.7)\times10^9$  M$_\odot$ an asymmetric bright emission ring with a diameter (shadow angular diameter) of $42 \pm 3$ $\mu as$, exhibiting a deviation from circularity $\Delta C \leq 0.1$ and an axis ratio $ 4/3$. These observational parameters is compatible with Kerr black hole in GR background. 

Additional observational data of the supermassive black hole Sgr A$^*$, published by the Event Horizon Telescope (EHT) \cite{akiyama2022firstv1}, revealed an angular shadow diameter $\theta_{sh}= 48.7 \pm 0.7$ $\mu as$ and Schwarzschild shadow deviation $-0.08^{+0.09}_{-0.09}$ (VLTI), $-0.04^{+0.09}_{-0.10}$ (Keck), with the black hole's mass $M = 4.0^{+1.1}_{-0.6} \times 10^6$ $M_\odot$ \cite{EventHorizonTelescope:2019ggy}.
\section{Theory background}
The black hole shadow and photon sphere can be calculated analytically in general theory of relativity or other theories. These depend on the metric tensor and geodesic for photon. 
Now, we have the Schwarzschild metric for static, non-rotating, and uncharged symmetric sphere \cite{1916SPAW:189S, narlikar2002introduction, schutz2022first}
\begin{equation}
	ds^2= -\Big(1 - \frac{2GM}{c^2r}\Big)c^2 dt^2 + \frac{dr^2}{\Big(1 - \frac{2GM}{c^2r}\Big)} +r^2(d\theta^2 + sin^2\theta d\phi^2)
	\label{EQ1}\end{equation}
The Lagrangian written as \cite{Weinberg:1972kfs}
\begin{equation}
	\mathcal{L} = \frac{1}{2}g_{\mu\nu}\dot x^\mu \dot x^\nu.
\end{equation}
If we consider the equatorial plane, where $(\theta =\pi/2$, $sin(\pi/2) = 1$ and $d\theta /d\lambda =0$ because of the photons trapped in the one plane and the path is circular.  The Lagrangian in the equatorial plane  is 
\begin{equation}
	\mathcal{L} = - \frac{1}{2} \Big(1- \frac{R_s}{r} \Big)\dot t^2 + \frac{1}{2}\frac{
		\dot r^2}{ \Big(1- \frac{R_s}{r} \Big)} +\frac{1}{2}r^2 \dot\phi^2,
\end{equation}
where  $R_s = 2GM/c^2$ is Schwarzschild radius of spherically symmetric object.

Euler-Lagrange equation is \cite{Weinberg:1972kfs, schutz2022first}

\begin{equation}
	\frac{d}{d\lambda}\Big(\frac{\partial \mathcal{L}}{\partial\dot x^\mu} \Big) - \frac{\partial\mathcal{L}}{\partial x^\mu} = 0.
	\label{EQ5}\end{equation}
Lagrangian have two cyclic coordinates $t$ and $\phi$ therefore we have two constant of motion ${\partial\mathcal{L}}/{\partial t}=0$ and ${\partial\mathcal{L}}/{\partial \phi}=0$. Because Lagrangian have two cyclic coordinates associated with two conserved quantities energy $E$ and angular momentum of photon $J_z$, respectively
\begin{equation}
	p_{t}=	\Big(\frac{\partial \mathcal{L}}{\partial\dot t} \Big)=\Big(1- \frac{2GM}{r} \Big)\dot t = E,
	\label{EQ9}\end{equation}
and 

\begin{equation}
	J_z =	\Big(\frac{\partial \mathcal{L}}{\partial\dot \phi} \Big)  =r^2 \dot \phi, 
	\label{EQ13}\end{equation}
where $c =1$. These are very useful quantities to study the trajectories of photon.
\subsection{Geodesic and photon orbit}
Einstein's general theory of relativity is most favorable theory for studying black hole \cite{bhlpage29122544}.    Imagine spacetime as a flexible fabric; massive objects like stars and planets create indentations, and black holes represent points where the fabric is stretched into an infinitesimal point-a singularity. 
We used Schwarzschild metric equation (\ref{EQ1}) to calculate photon sphere.

Now, we have calculated  the mathematical and geometrical expression of the photon sphere. The Photon moves around the massive object like black hole on the null geodesic i.e. $ds^2 = 0$ or $g_{\mu\nu}dx^\mu dx^\nu = 0$.   

Null geodesic equation in the equatorial plane is \cite{Zeng:2021dlj}
\begin{equation}
	- \Big(1- \frac{R_s}{r} \Big)dt^2 + \frac{dr^2}{\Big(1- \frac{R_s}{r} \Big)} + r^2 d\phi^2 = 0. 
\end{equation}
If we divide by the affine parameter $d\lambda$ \cite{Zeng:2021dlj}
\begin{equation}
	- \Big(1- \frac{R_s}{r} \Big)\dot t^2 + \frac{\dot r^2}{\Big(1- \frac{R_s}{r} \Big)} + r^2 \dot\phi^2 = 0. 
	\label{EQ16}\end{equation}
Above equation (\ref{EQ16}) used to calculate the size of the photon sphere and effective potential.

\begin{figure}[h]
	\centering
	\includegraphics[width=0.40\textwidth]{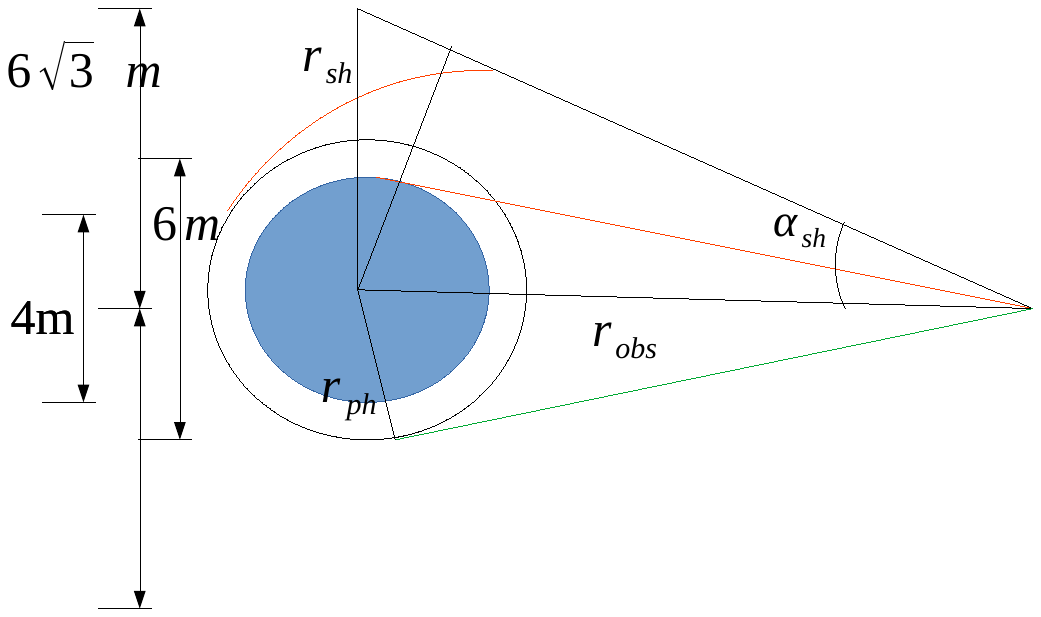}
	\caption{The dotted line is known as the angular diameter of black hole shadow $\alpha_{sh}$ or acceptance angle $ r_{sh} = 3\sqrt{3}m$ and red circle is the photon sphere $r_{ph} = 3m$ for static observer. If observer is at the large distance $r_o>>m$, then $r_{sh}$ equal to impact factor \cite{Wang2022kvg, Perlick2021CalculatingBH}. This figure shows that the sizes of black hole horizon, photon sphere and black hole shadow, respectively in Schwarzschild spacetime metric and for static observer}
	\label{fig:mesh0}
\end{figure}


\subsection{Effective Potential}
The effective potential can be used to calculate the radius of the photon sphere by $dV_{eff}/dr = 0$ \cite{Zeng:2021dlj}. Now calculate the effective potential using the null geodesic equation (\ref{EQ16}) and using the value of $E$ $\&$ $J_z$ from equations (\ref{EQ9}) and \ref{EQ13}, respectively.
\begin{equation}
	- \frac{E^2}{\Big(1- \frac{R_s}{r} \Big)}+ \frac{\dot r^2}{\Big(1- \frac{R_s}{r} \Big)} + \frac{J^2}{r^2} = 0, 
\end{equation}

\begin{equation}
	\frac{1}{2}\Big(\frac{dr}{d\lambda} \Big)^2 + \frac{1}{2}\frac{J^2}{r^2}\Big(1- \frac{R_s}{r} \Big) = E, 
	\label{EQ19}\end{equation}
where the effective potential $V_{eff}$ is 
\begin{equation}
	V_{eff} = \frac{J^2}{2r^2} -\frac{R_s}{2r^3}J^2.
	\label{EQ20}\end{equation}
The effective potential used to visualize the orbital motion of the photon. 
Equation (\ref{EQ19}) can be written as $K.E. + V_{eff} = E$.
The first term and second of the equation (\ref{EQ20}) are the centrifugal potential and the gravitational potential, respectively. 

\subsection{Photon Sphere }
Radius of the photon sphere calculated from null geodesic, i.e., the geodesic for photon in the equatorial plane. 
The radius of the photon sphere calculated by using effective potential (\ref{EQ20}) and differentiating with respect to $r$ \cite{Wang:2022kvg, Perlick2021CalculatingBH}
\begin{equation}
	\frac{dV_{eff}}{dr} = -\frac{J^2}{r^3} +\frac{3}{2}\frac{J^2}{r^4} R_s
\end{equation}
Value of $r$ at least point of $V_{eff}$ is 
\begin{equation}
	-\frac{J^2}{r^3} +\frac{3}{2}\frac{J^2}{r^4}R_s = 0
\end{equation}
\begin{equation}
	r_{ph} = \frac{3}{2}R_s=3m,
\end{equation}
where, $R_s = G_N M/c^2$,  $m = G_N M/c^2$ and $m$ has the dimension of the length. The
value of the second derivative of the $V_{eff}$ tells about the minimum (stable point) and maximum (unstable point) of the $V_{eff}$. Now we obtained  $V_{eff}$ has minima or maxima as

\begin{equation}
	\frac{d^2V_{eff}}{dr^2} = \frac{3J^2}{r^4} -\frac{6J^2}{r^5}R_s.
	\label{EQ24}\end{equation}
After putting $r= (3/2) R_s$, then
\begin{equation}
	\frac{d^2V_{eff}}{dr^2} = \frac{16}{27}\frac{J^2}{R_s^4} -\frac{64}{81}\frac{J^2}{R_s^5}= -\frac{16}{27}\frac{J^2}{R_s^4}.
\end{equation}
At $r_{ph} = (3/2) R_s$, potential has unstable point because $d^2V_{eff}/dr^2<0$. 
\subsection{Shadow of Black hole }
Mathematical explanation of the black hole shadow for the first time given by Synge \cite{synge1966escape} in 1996. He considered spherical symmetric Schwarzschild metric foe static universe. He takes $\rho = r/2m$ and $\tau = t/2m$, where $G=c=1$. Constant of motion for angular momentum and energy conservation given as $\rho^2\dot\theta = \alpha$ and $(1-\rho^{-1})\dot t = \alpha\beta$, respectively. Path of the photon $\theta=$ const. $\&$ $\phi = $const. and Small deviation of light ray (photon) is $\cot \alpha = \sqrt{g_{\rho\rho}/g_{\theta\theta}}(d\rho/d\theta)$ and $\sin^2 \alpha = (\rho -1)/\rho^3\beta^2$. We have  $\beta= 4/27$ for $\rho=3/2$. 

The Synge equation is 
\begin{equation}
	\sin^2\alpha_{sh} = \frac{27}{4}\frac{\rho_o -1}{\rho_o^3}.
	\label{EQ26}\end{equation}
Equation (\ref{EQ26}) gives the angular radius of the black hole shadow.
\section{General expression of Black Hole Shadow and photon sphere}
Most of the part of this section use the review article \cite{Wang:2022kvg, Perlick2021CalculatingBH}. We have find out the general solution of the black hole shadow for spherical symmetric case in the static universe. In this case metric is 
\begin{equation}
	ds^2 = -A(r) dt^2 + B(r) dr^2 + D(r) (d\theta^2 +\sin^2\theta d\phi^2), 
	\label{IVA1}\end{equation}
where, $A(r)$, $B(r)$ and $D(r)$ are constant. If we use $A(r) = 1- 2m/r$, $B(r) = (1- 2m/r)^{-1} $ and $D(r) = r^2$, $m = G_N M/c^2$.
We find constant of motion equation form (\ref{EQ5}) as
\begin{align}
	-p_t = E = -A(r) \dot t, &&  p_\theta =J_z = D(r) \sin^2\theta\dot\phi 
\end{align}
where $E$, and $J_z$ are energy and angular momentum in z-direction. We define the impact parameter as $b\equiv J_z/E$. If light is in equatorial plane $\theta = \pi/2$ then
\begin{align}
	E = -A(r) \dot t, &&  J_z = D(r)\dot\phi.
	\label{IV23}\end{align}

We have effective potential $V_{eff}$ from Hamiltonian \cite{Wang2022kvg}, $\mathcal{H} = (1/2) g^{\mu\nu} p_\mu p_\nu $
\begin{equation}
	\mathcal{H} = \frac{1}{2}\Big(\frac{p_t^2}{A(r)}+\frac{p_r^2}{B(r)} + \frac{p_\theta^2}{D(r)} +\frac{p_\phi^2}{D(r)\sin^2\theta}\Big),,
\end{equation}
\begin{equation}
	\mathcal{H} = \frac{1}{2}\Big(-\frac{E^2}{A(r)}+\frac{p_r^2}{B(r)} + \frac{p_\theta^2}{D(r)} +\frac{J_z^2}{D(r)\sin^2\theta}\Big)
\end{equation}

\begin{equation}
	\mathcal{H} = \frac{1}{2}\Big(\frac{p_r^2}{B(r)} + \frac{p_\theta^2}{D(r)} + V_{eff} \Big),
\end{equation}
where $ V_{eff}$ \cite{Wang2022kvg}, 
\begin{equation}
	V_{eff} = -\frac{E^2}{A(r)} +\frac{J_z^2}{D(r)\sin^2\theta}.
\end{equation}
If, we put the value of $A(r)$ , $B(r)$, $D(r)$ and $\theta = \pi/2$ in equatorial plane so we get equation \eqref{EQ20} of effective potential. There are two conditions for circular motion of photon as $V_{eff} = 0$ and $\partial V_{eff}/\partial r = 0$ should be follow.

Geodesic equation for photon $g_{\mu\nu} \dot x^\mu \dot x^\nu = 0$ is  \cite{ Perlick2021CalculatingBH}
\begin{equation}
	-A(r) \dot t^2 + B(r) \dot r^2 + D(r) \dot\phi^2 = 0.
	\label{IV28}\end{equation}
Now using equation \eqref{IV23} and \eqref{IV28}, we get $\dot r^2/ \dot t^2 = (dr/d\phi)$ \cite{ Perlick2021CalculatingBH}
\begin{equation}
	\Big(\frac{dr}{d\phi}\Big)^2 = \frac{D(r)}{B(r)}\Big(\frac{D(r)}{A(r)}\frac{1}{b^2}-1  \Big).
	\label{IV29}\end{equation}
From the above equation, we can deduced impact parameter $b$ by imposing the condition $dr/d\phi|_R = 0$  \cite{ Perlick2021CalculatingBH}.
\begin{equation}
	b^2 = \frac{E^2}{J_z^2}=\frac{D(R)}{A(R)}.
	\label{IV30}\end{equation}
Now, we define the function \cite{ Perlick2021CalculatingBH}
\begin{equation}
	h^2(x) \equiv \frac{D(x)}{A(x)}
	\label{IV31}\end{equation}
Angular radius of the shadow is angle between light ray and radial direction for static observer at $r=r_o$
\begin{equation}
	\cot\alpha = \frac{\sqrt{g_{rr}}}{\sqrt{g_{\phi\phi}}}\frac{dr}{d\phi}=\frac{\sqrt{B(r)}}{\sqrt{D(r)}}\frac{dr}{d\phi}.
\end{equation}
After using equation \eqref{IV29} and \eqref{IV31}, we obtain
\begin{equation}
	\cot^2\alpha = \frac{h^2(r)}{h^2(R)} -1,
\end{equation}
and 
\begin{equation}
	\sin^2\alpha = \frac{h^2(R)}{h^2(r)}.
\end{equation}
When $R$ becomes radius of photon sphere $r_{ph}$ \cite{ Perlick2021CalculatingBH} 
\begin{equation}
	\sin^2\alpha_{sh} = \frac{h^2(r_{ph})}{h^2(r)}.
\end{equation}
The critical impact parameter $r_{sh}\equiv b_{cr}$ is equal to $r_{ph}/\sqrt{A(r_{ph})}$. This represented in the fig. \eqref{fig:mesh0}.

The generalized condition for radius of the photon sphere is
\begin{eqnarray}
	\frac{dh^2(r)}{dr} = 0.
	\label{IV35}\end{eqnarray}

This equation is useful to calculate the radius of the photon sphere in GR and $f(R)$ gravity.

\subsection{Shadow in Schwarzschild space-time for static observer}
Here we have the parameter $A(r), B(r), D(r)$ in the Schwarzschild metric \eqref{EQ1} as \cite{Perlick2021CalculatingBH}
\begin{align}
	A(r) = 1- \frac{2m}{r}, && B(r) = \Big(1- \frac{2m}{r}  \Big)^{-1}, && D(r) = r^2,
\end{align}
where, $m = G_N M/c^2$ is not a mass of the black hole. It has the dimension of the length.
We  have the function $h^2(r) ={\frac{D(r)}{A(r)}}= \frac{r^2}{1-2m/r}$ and
obtained the radius of the photon sphere by using equation \eqref{IV35} is 
\begin{equation}
	\frac{d}{dr}\Bigg(\frac{r^2}{1-2m/r} \Bigg) = \frac{3(r-2m)r^2 -r^3}{(r-2m)^2} ,
\end{equation}
\begin{align}
	\frac{dh^2(r)}{dr} = 0, && \frac{3(r-2m)r^2 -r^3}{(r-2m)^2} = 0, && r_{ph}= 3m.
\end{align}
The critical impact parameter is $b_{cr} =h(r_{ph})$ ,
\begin{equation}
	b^2_{cr} = \frac{r^2_{ph}}{1 -2m/r_{ph}} = \frac{9m^2}{1 -2m/3m}= 27m^2,
\end{equation}
\begin{equation}
	b^2_{cr} = 3\sqrt{3} m.
\end{equation}
In the fig. \eqref{fig:mesh0}, we can distinguished between event horizon ($2m$), photon sphere ($3m$) and black hole shadow ($3\sqrt{3}$) in Schwarzschild spacetime.  Angular radius of the black hole shadow for large distance $r_o >> m$ is
\begin{equation}
	\alpha_{sh} = \frac{3\sqrt{3}m}{r_o}.
	\label{IVA41}\end{equation}
The equation \eqref{IVA41} is useful to calculate angular diameter of the black hole shadow.

\subsection{Shadow in the Kottler space-time for a static observer}
Here we discussed the effect of the cosmological constant on BHS and photon sphere. We considered the Kottler metric \cite{Kottler1918A,Perlick2021CalculatingBH}
\begin{align}
	A(r)= 1-\frac{2m}{r} -\frac{\Lambda}{3}r^2,  && B(r) = \Bigg( 1-\frac{2m}{r} -\frac{\Lambda}{3}r^2 \Bigg)^{-1},  && D(r) = r^2.
\end{align}
We have calculated photon sphere $r_{ph}$ and critical impact parameter $b_{cr}$ by using $dh^2{r}/dr = 0$
\begin{equation}
	h^2(r) = \frac{r^2}{1-2m/r -\Lambda r^2/3},
\end{equation}

\begin{equation}
	\frac{dh^2(r)}{dr}= \frac{(3r - 6m -\Lambda r^2)3dr^2/dr - 3r^2d(3r-6m-\Lambda r^3)/dr}{3r^2}=0.
\end{equation}
Photon sphere $r_{ph}= 3m $ and critical impact parameter $b_{cr} = 3\sqrt{3}/\sqrt{(1-9\Lambda m^2)}$. Now, we can calculated angular radius of the black hole shadow is
\begin{equation}
	\sin^2{\alpha_{sh}} = \frac{h^2(R)}{h^2(r_o)} = \frac{b^2_{cr}}{h^2(r_{o})},
\end{equation}
\begin{equation}
	\sin^2{\alpha_{sh}} = \frac{27m^2(1-2m/r_o - \Lambda r^2_o/3)}{(1-9\Lambda m^2) r^2_o}.
	\label{VIB46}\end{equation}
Here we find two horizon \cite{Wang2022kvg, Perlick2021CalculatingBH} for $0< \Lambda < (3m)^{-2}$ and also find out that the ratio of impact parameter and observer distance does not equal to angular radius of the black hole shadow for static observer just like in Schwarzschild space-time. Therefore we considered commoving observer with cosmological accelerated expansion.
\subsection{Shadow of Kerr black hole}
Kerr black hole is an uncharged black hole with rotational motion. We have discussed the space-time around the black hole in vacuum with the help of Kerr metric \cite{Kerr1963ud, Boyer1966qh,Perlick2021CalculatingBH} is 
\begin{widetext}
	\begin{equation}
		ds^2 = - \Bigg(1 -\frac{2Mr}{\rho^2} \Bigg)dt^2 + \frac{\rho^2}{\Delta} dr^2 + \rho^2 d\theta^2 + \sin^2\theta\Bigg( r^2 + a^2 + \frac{2Mra \sin^2\theta}{\rho^2}\Bigg) d\phi^2 -\frac{4M ra \sin^2\theta}{\rho^2} dtd\theta,
	\end{equation}
	where
	\begin{align}
		\Delta = a^2 + r^2 -2Mr, && \rho^2 = r^2 +a^2 \cos^2\theta, 
	\end{align}
	$M$ and $a$ is the mass and spin parameter of the Kerr Black bole. The Hamiltonian $ H(x,p) = \frac{1}{2} g^{\mu\nu}(x)p_\mu p_\nu$ of null geodesic is 
	\begin{equation}
		H(x,p) = \frac{\Delta}{2\rho^2} P_r^2 + \frac{1}{2\Delta\rho^2}[(r^2 + a^2)P_t + a P_\phi]^2  
		+\frac{1}{2\rho^2}P_\theta + \frac{1}{2\rho^2\sin^2\theta}(P_\phi +a P_t \sin^2\theta)^2.
	\end{equation}
\end{widetext}
Integration of the geodesic equation gives three constants in the Kerr black hole, $E$ energy, $z$-component of the angular momentum $J_z$ 
\begin{align}
	E = -p_t= -g_{tt} \dot t -g_{t\phi}\dot\phi,  && \\
	J_z= p_\phi = g_{\phi\phi}\dot\phi + g_{\phi t} \dot t.
\end{align}
We have the null geodesic equations
\begin{align}
	\dot t = E + \frac{2Mr(a^2 E -aJ_z + Er^2)}{\Delta\rho^2}&& \\
	\dot\phi = \frac{2MraE}{\Delta\rho^2 } +  \frac{\Delta - a^2\sin^2\theta}{\Delta\rho^2 \sin^2\theta}J_z, && \\
	R(r)= \rho^4 \dot r^2 =\Delta^2 p^2_r  \nonumber && \\ = -\Delta[Q + (aE -J_z)^2] + [aJ_z -(r^2 +a^2)E]^2,&& \\
	\Theta(\theta) = \rho^4\dot\theta^2=p^2_\theta = Q -\cos^2\theta\Bigg(\frac{J^2_z}{\sin^2\theta} - a^2E^2 \Bigg).
\end{align}
We find out the $R^\prime(r)$ after differentiating with respect to $r$ is 
\begin{align}
	R^\prime(r) = -4Er[aJ_z - (r^2 +a^2)E] \nonumber && \\ 
	-2(r-M)[Q + (aE -J_z)^2].    
\end{align}

After using boundary condition $R^\prime(r) = 0$, we have impact parameter $\eta = J_z/E$ and  constant $\sigma =Q/E^2$ for spherical orbit of photon is
\begin{equation}
	\eta = -\frac{r^2(r-3M) + a^2(r+M)}{a(r-M)};
\end{equation}
\begin{equation}
	\sigma=  \frac{r^3[4a^2M - r(r-3M)^2]}{a^2(r-M)^2}.
\end{equation}
Using above parameters, we have calculated the coordinates$(x,y)$ of a point in the observer plane and static observer is locally at $(r_o, \theta_o)$ in zero-angular momentum frame of reference \cite{1973Bardeen, Wang2022kvg, Johannsen2013vgc}. If the observer is at infinity $(r_o \rightarrow \infty)$ means far away from the black hole. We have \cite{Wang2022kvg}
\begin{align}
	x = -\frac{\eta}{\sin\theta_o}; && \nonumber \\ 
	y =\pm \sqrt{\sigma +a^2\cos^2\theta_o - \eta^2 \cot^2\theta_o}.
\end{align}
We find out the shape of the Kerr black hole shadow with the help of above equations $x$ and $y$. The unstable prograde and retrograde photon rings gives 
\begin{equation}
	r_{lr} = 2M\Bigg(1+ \cos[(2/3) \arccos (\pm a/M)]  \Bigg),
\end{equation}
where $-$ and $+$ denote prograde and retrograde respectively. The deviation parameter is 
\begin{equation}
	\delta = \frac{D}{R_{sh}},
\end{equation}
where $D$ is the horizontal displacement of the photon ring with respect  to the
center of the image plane by the expression
\begin{equation}
	D = \frac{x_{min} + x_{max}}{2},
\end{equation}
where $x_{max}$ and $x_{min}$ are the maximum and minimum abscissae, respectively
of the ring, respectively
This is the deviation of black hole shadow from circle. The average radius of the photon ring is \cite{Johannsen2013vgc}
\begin{equation}
	\langle \bar R \rangle = \frac{1}{2\pi} \int_0^{2\pi} \bar R d\alpha,
\end{equation}
where $\bar R = \sqrt{(x-D)^2 + y^2}$ and $\tan\alpha = y/x$. We can used deviation parameter $\delta$, horizontal displacement $D$ and average radius $\langle \bar R \rangle$ for constraint the parameters of $f(R)$ gravity model and other theories.

\section{Black hole shadow and Photon sphere in $f(R)$ gravity}
The modified theory of gravity explain unsolved problems elegantly \cite{Sharma2022fiw, Sharma2022tce, Sharma2019yix, KumarSharma2022qdf}. In this article, we have used $f(R)$ background in place of GR background for the black hole shadow and photon sphere. In 
$f(R)$ gravity, modifications to the spacetime curvature can alter trajectories of photon, changing the size and shape of the shadow \cite{nojiri2024black, nojiri2025black}. These deviations appear because of the specific form of $f(R)$, the black hole's spin, and surrounding matter fields.

\subsection{Field equations in $f(R)$ gravity}
We considered static and spherical symmetric spacetime metric \cite{capozziello2019maximum} given by equation  \eqref{IVA1}, 
\begin{equation}
	A(r) = B(r)^{-1} = 1 - \frac{2G_{eff} M}{r} -\frac{f(R)}{6f^\prime(R)}r^2\Bigg|_{R_{dS}},
	\label{VIA1}\end{equation}
where $G_{eff} = G_N/F$, and $F = f^\prime(R)=df(R)/dR$.
Einstein-Hilbert action in $f(R)$ gravity is \cite{de2010f, nojiri2017modified}
\begin{equation}
	\mathcal{A} = \int{d^4x\sqrt{-g}}\Bigg(\frac{1}{16\pi G_N} f(R) + \mathcal{L}_m \Bigg).
\end{equation}
After taking variation with respect to the metric, we have 
\begin{equation}
	f^\prime(R) R_{\mu\nu} -\frac{1}{2} f(R) g_{\mu\nu} + (g_{\mu\nu}\Box -\nabla_\mu\nabla_\nu)f^\prime(R) = 8\pi G_NT^M_{\mu\nu},
\end{equation}
where $T^M_{\mu\nu}$ is the energy momentum tensor of matter. The trace of the above equation is \cite{de:2010f,nojiri:2017modified}
\begin{equation}
	f^\prime(R) R -2 f(R) + 3\Box f^\prime(R) = 8\pi G_NT^M.
\end{equation}
For spherically symmetric case, we used constant curvature solution of $f(R)$ model i.e. $R=R_{dS} =constant$. The equation \eqref{VIA1}  becomes  \cite{capozziello2019maximum} 
\begin{equation}
	A(r) = B(r)^{-1} = 1 - \frac{2G_{eff} M}{r} -\frac{R_{dS}}{12}r^2.
	\label{VIA79}\end{equation}
We can easily get the Lagrangian $\mathcal{L} = (1/2)g_{\mu\nu}\dot{x^\mu}\dot{x^\nu}$ from the equation \eqref{IVA1} 
\begin{equation}
	\mathcal{L} = \frac{1}{2} [-A(r) \dot{t}^2 + B(r)\dot{r}^2 + r^2 \dot{\theta}^2 +r^2 \sin^2\theta\dot\phi^2].
\end{equation}
We have calculated four momentum as
\begin{align}
	-p_t=\frac{\partial\mathcal{L}}{\partial\dot t} = E  = -A(r)\dot t; &&  p_\phi = \frac{\partial\mathcal{L}}{\partial\dot \phi}= r^2\sin^2\theta \dot\phi \\
	p_\theta = r^2\dot\theta;  && p_r = B(r) \dot r.   
\end{align}
Null geodesic equation is 
\begin{equation}
	-\frac{E^2}{A(r)} + B(r) \dot r^2 + \frac{J^2}{r^2} = 0.
	\label{VA71}\end{equation}
Using the equation \eqref{VIA79} in the equation \eqref{VA71}, we have
\begin{equation}
	\frac{1}{2} \dot r^2 + \frac{J^2}{r^2} \Bigg(1 -\frac{2G_{eff}M}{r} - \frac{R_{dS}}{12} r^2 \Bigg) =
	\frac{E^2}{2}.
	\label{VA72}\end{equation}
The effective potential obtained from equation \eqref{VA72}
\begin{equation}
	V_{eff} = \frac{1}{2}\frac{J^2}{r^2} \Bigg(1 -\frac{2G_{eff}M}{r} - \frac{R_{dS}}{12} r^2 \Bigg).
\end{equation}
We have calculated the radius of the photon sphere by using $dV_{eff}/dr = 0$, is
\begin{equation}
	\frac{J^2}{r^3} - \frac{3}{2}\frac{G_{eff}M J^2}{r^4} = 0,
\end{equation}
\begin{equation}
	r_{ph} = \frac{3}{2}{G_{eff}M } = \frac{3}{2}\frac{G_{N}M }{F} = \frac{3}{2} \frac{r_s}{F}.
	\label{VIA85}\end{equation}
Radius of the photon sphere in the $f(R)$ gravity larger than  in the  GR background because of modification. We can recovered $r_{ph}=(3/2)r_s$, if $F =1$.

Next, we have calculated black hole shadow and impact factor in $f(R)$ background. We used $dh^2(r)/dr = 0$, where
\begin{equation}
	h^2(R) = \frac{r^2}{1 - \frac{2G_{eff} M}{r} -\frac{R_{dS}}{12}r^2},
\end{equation}
and 
\begin{equation}
	\frac{dh^2(R)}{dr} = 2r\Bigg(1 - \frac{2G_{eff} M}{r} -\frac{R_{dS}}{12}r^2  \Bigg) -r^2\Bigg( \frac{2G_{eff}M}{r} - \frac{rR_{dS}}{6}\Bigg) =0,
\end{equation}
\begin{equation}
	2r - 4G_{eff}M - \frac{r^3R_{dS }}{6} -2G_{eff}M +\frac{r^3R_{dS }}{6} =0,
\end{equation}
\begin{equation}
	r_{ph} = 3G_{eff}M = \frac{3G_N M}{F}.
	\label{VIA91}\end{equation}
If we considered $F =1$, radius of the photon sphere reduce in form $3G_N M$, which is same as in GR background. The photon sphere change because of modification in the background. We have the critical impact factor $b_{cr}$ by using the equation \eqref{IV30} is
\begin{equation}
	b^2_{cr} = \frac{r_{ph}^2}{1 - \frac{2G_{eff} M}{r_{ph}} -\frac{R_{dS}}{12}r_{ph}^2}.
	\label{VIA92}\end{equation}
After putting the value of $r_{ph}$ from the equation \eqref{VIA91} in equation \eqref{VIA92}, we have
\begin{equation}
	b^2_{cr} = \frac{9G_{eff}^2M^2}{\frac{1}{3}- \frac{3}{4}G_{eff}^2M^2R_{dS}},
\end{equation}

\begin{equation}
	b_{cr} = \frac{3\sqrt{3}G_{eff}M}{\sqrt{1- \frac{9}{4}G_{eff}^2M^2R_{dS}}} = \frac{3\sqrt{3}G_{eff}M}{\sqrt{1- 2.25 G_{eff}^2M^2R_{dS}}},
\end{equation}

\begin{equation}
	b_{cr} = \frac{3\sqrt{3}G_{N}M/F}{\sqrt{1- 2.25R_{dS} G_{eff}^2M^2/F^2}}.
\end{equation}
Now, we have calculated black hole shadow as
\begin{equation}
	\sin^2\alpha_{sh} = \frac{b_{cr}^2}{h^2(r_o)},
\end{equation}
\begin{equation}
	\sin^2\alpha_{sh} = \frac{9G_{eff}^2M^2\Big(1 - \frac{2G_{eff} M}{r_{o}} -\frac{R_{dS}}{12}r_{o}^2\Big)}{r_o^2(1- 2.25 G_{eff}^2M^2R_{dS})},
\end{equation}
\begin{equation}
	\sin^2\alpha_{sh} =\Big(1 - \frac{2G_{eff} M}{r_{o}} -\frac{R_{dS}}{12}r_{o}^2\Big) \frac{b^2_{cr}}{r_o^2},
	\label{VI92}\end{equation}
\begin{equation}
	\sin^2\alpha_{sh} =\Big(1 - \frac{2G_{N} M}{r_{o}F} -\frac{R_{dS}}{12}r_{o}^2\Big) \frac{b^2_{cr}}{r_o^2}.
	\label{VI93}\end{equation}

The above equation \eqref{VI93}  of black hole shadow look similar to the equation \eqref{VIB46} in Kottler spcetime. There are two changes appeared in the equation \eqref{VI93}; the last term behaves as the $\Lambda$ and extra scalar degree of freedom $F$ appeared in the second term. These factor appeared in the expression because of the modification in curvature. 

\section{Conclusion}
In this paper, we have found that the size of the black hole photon sphere and shadow changed with metric tensor and new scalar degree fo freedom $F$ appears in the  modified theory of gravity gravity. Indeed, There is no need to introduced two components separately in the modified gravity. Therefore, we preferred $f(R)$ theory of gravity. We obtained generalized formulae of photon sphere (\ref{VIA91}) and black hole shadow (\ref{VI92}) for $f(R)$ theory.

In Kottler spacetime, there is two horizon and we can also find two horizon in $f(R)$ gravity from the equation (\ref{VI92}). If we considered the value $F= 1$ and $R_{dS}=4\Lambda$, we got similar expression of  black hole shadow in Kottler spacetime. That means, we recovered the expression in GR background.

The size of the photon sphere behave inversely with $F$ and BHS also affected through the modification in gravity. These effects shown through the equation \eqref{VIA91} or \eqref{VIA85}. These are our main finding of this paper. Black hole shadow and photon sphere changed in modified theory of gravity.

In future, we will explore the effect of the modified gravity on the accretion disc. The distortion of shadow needs to be explained through the viable theory of gravity.  We will also use Chameleon mechanism in high density regime and search for dark matter candidate through the black hole shadow.

\begin{center}
\textit{\underline{Acknowledgments}}
\end{center}
 Authors are thankful to  Vipin Sharma for the useful   discussions  on  various aspects of  black hole.

\bibliography{Shadow1}
\end{document}